\documentclass[twoside,a4paper,10pt]{article}
\usepackage[latin1]{inputenc}
\usepackage[T1]{fontenc}
\usepackage{amsmath}
\usepackage{amsfonts}
\usepackage{graphicx}
\usepackage{subfigure}
\usepackage{a4wide}
\usepackage{amssymb}
\usepackage{fancyhdr}
\usepackage{mathrsfs}

\def\beq{\begin{eqnarray}}
\def\eeq{\end{eqnarray}}

\def\be{\begin{eqnarray}}
\def\ee{\end{eqnarray}}

\begin{document}

\title{\bf On the Energy Issue for a Class of  Modified Higher Order Gravity Black Hole Solutions  }
\author{ 
G. Cognola \footnote{E-mail address:cognola@science.unitn.it}\,,O. Gorbunova \footnote{olesya@science.unitn.it}\,,L. Sebastiani\footnote{E-mail address:l.sebastiani@science.unitn.it
}\, and S. Zerbini\footnote{E-mail address:zerbini@science.unitn.it
}\\
\\
\small
Dipartimento di Fisica, Universit\`a di Trento
and Istituto Nazionale di Fisica Nucleare,\\ 
\small 
Sezione di Padova, Gruppo Collegato di Trento, Italia
}
\maketitle 


\begin{abstract}
In the case of a large class of static, spherically symmetric black hole solutions 
in higher order modified gravity models,
 an expression for the  associated energy is proposed and identified with
a quantity proportional to the constant of integration, which appears 
in the explicit solution. 
The identification is achieved making use of 
derivation of the First Law of black hole thermodynamics from the equations of motion, 
evaluating independently the entropy via Wald method and the 
Hawking temperature via quantum mechanical methods in curved space-times. 
Several non trivial examples are discussed, including a new topological higher 
derivative black hole solution, and the proposal is shown to work in all examples
considered. 

\end{abstract}

PACS: {04.50.Kd; 04.70.Dy; 97.60.Lf; 95.30.Sf}


\section{Introduction}

Recent observational data imply an accelerating expansion of the visible universe, 
which gives rise to the so called Dark Energy issue.

There exist several descriptions of  this acceleration. 
Among them, the simplest one consists in the introduction of a small positive 
cosmological constant in the framework of General Relativity (GR), 
the so called $\Lambda$-CDM model. 
A generalization of this simple modification of GR consists in considering  
modified gravitational theories, 
in which the action is described by a function $F(R)$ of the Ricci scalar $R$ 
(see for example \cite{review6,review7}). 
Typically these modified models admit the de Sitter space as 
a solution and the stability of this solution has been investigated in several places  
(see for example \cite{guido,guido2,Far,Monica}). Furthermore, viable $F(R)$ models, 
that is the ones which are able to pass the local gravitational GR tests, 
as well as to describe the inflation with dark energy in a unified way, 
have been recently discussed \cite{Saw,seba10,Od1,Od2}. 
Another very interesting class of modified gravitational models 
in which the square root  of the quadratic Weyl 
scalar appears have been investigated in Ref. \cite{Deser:2007za}.

Static, spherically symmetric solutions have been investigated in several papers 
(the simplest one being the Schwarzschild-de Sitter solution), 
and they have been discussed for example in Refs.\cite{CB,Zerbini,altri,Deser:2007za}.
Within this class of higher order gravitational models, 
the issue associated with the energy (mass) of black hole solutions is problematic, 
and several attempts in order to find a satisfactory answer to that problem
have been proposed (see for example \cite{Do,D,V1,Cai} and references therein). 

To start with, let us remind the case of GR, in which several notions of quasi-local 
energies may be introduced. 
In particular we mention the so called Misner-Sharp mass, 
which has the important property to be  defined for dynamical, 
spherically symmetric space-time \cite{H}, where 
the use of invariant quantities plays a crucial role \cite{noi2,noi3}.  
For the sake of completeness, we recall that in four dimensions, 
any spherically symmetric metric can locally be expressed in the form   
\beq
\label{metric1}
ds^2 =\gamma_{ij}(x^i)dx^idx^j+ {\mathcal R}^2(x^i) d\Omega_2^2\,,\qquad i,j \in \{0,1\}\;,
\eeq
where $ d\Omega_2^2 $ here is the usual metric on the two sphere $S^2$, but it could be
be the metric of a generic two-dimensional maximally symmetric space. 
Of course, in such cases the black hole will have a different topology. 
The two-dimensional metric
\beq d\gamma^2=\gamma_{ij}(x^i)dx^idx^j
\label{nm}
\eeq
is referred to as the normal one. The related coordinates are $\{x^i\}$, while
$ {\mathcal R}(x^i)$ is the areal radius, considered as a scalar field in the two dimensional
normal space. 
A relevant scalar quantity in the reduced normal space is 
\beq
\chi(x)=\gamma^{ij}(x)\partial_i  {\mathcal R}(x)\partial_j  {\mathcal R}(x)\,, \label{sh} 
\eeq 
since the dynamical trapping horizon, if it exists, is located in
correspondence of 
\beq 
\chi(x)\Big\vert_H = 0\,, \label{ho} 
\eeq
provided that $\partial_i\chi\vert_H \neq 0$.
(We use the suffix ${}\vert_H$ for all quantities evaluated on the horizon).
The quasi-local Misner-Sharp gravitational energy is defined by
\beq
E_{MS}(x)=\frac{1}{2}  {\mathcal R}(x)\left[1-\chi(x) \right]\,. \label{MS}
\eeq
This is an invariant quantity on the normal space. Note also that, on
the horizon, $E_{MS}\vert_H =\frac{1}{2}  {\mathcal R}_H\equiv E$,
$E$ being the energy of black hole. Furthermore, one can
introduce the Hayward surface gravity associated with this dynamical
horizon, which is given by the normal-space scalar 
\beq
\kappa_H=\frac{1}{2}\Box_{\gamma} {\mathcal R}\Big\vert_H\,, \label{H} 
\eeq 
$\Box_\gamma$ being the Laplacian corresponding to the $\gamma$ metric.
In the spherical symmetric, dynamical case, it is also possible to introduce 
the Kodama vector field $\mathcal K$. 
Given the metric (\ref{metric1}) it is defined by
\beq 
\mathcal K^i(x)=\frac{1}{ \sqrt{-\gamma}}\,\varepsilon^{ij}\partial_j  {\mathcal R}\,,
\qquad \mathcal K^\theta=0=\mathcal K^\varphi \label{ko} \;,
\eeq 
$\varepsilon^{ij}$ being the completely antisymmetric Levi-Civita tensor on the normal 
space.

Assuming Einstein equations, in a generic four-dimensional 
spherically symmetric space-time, a geometric dynamical identity holds true in general. 
This can be derived as follows. 
Let us introduce the normal space invariant 
\beq
T^{(2)}=\gamma^{ij}T_{ij}\,,
\eeq 
which is the reduced trace of the stress energy tensor $T_{\mu\nu}$. 
Then, making use of Einstein equations, it is possible to show that, 
on the dynamical horizon (see for example\cite{H})
\beq
\kappa_H=\frac{1}{2 {\mathcal R}_H}+2\pi  {\mathcal R}_H T^{(2)}_H\,.
\label{vanzo}
\eeq
Introducing the horizon area $\mathcal A_H$ 
and the (formal) three-volume $V_H$ enclosed by the horizon, 
with their respective ``thermodynamical'' differentials 
$d\mathcal A_H=8\pi  {\mathcal R}_H d {\mathcal R}_H\,$, 
and $d V_H=4\pi  {\mathcal R}_H^2 d {\mathcal R}_H$
(we are assuming a horizon with the topology of a sphere), we get
\beq
\frac{\kappa_{H}}{8 \pi}d \mathcal A_H =d\left(\frac{ {\mathcal R}_H}{2}\right) 
+\frac{ T_H^{(2)}}{2} dV_H\,. 
\eeq
This equation can be recast in the form of a geometrical identity, 
once the Misner-Sharp energy at the horizon has been introduced. It reads 
\beq
dE=\frac{\kappa_{H}}{2 \pi} d\left( \frac{\mathcal A_H}{4}\right)
-\frac{T_H^{(2)}}{2} dV_H\,. 
\label{fl}
\eeq

In the following, we shall  restrict the discussion to the static case
in the absence of matter. This means that we shall consider only vacuum static 
solutions. In such a case the metric in (\ref{metric1}) can be written in the simpler form 
\begin{eqnarray}
ds^2=-B(r)e^{2\alpha(r)}dt^2+\frac{dr^2}{B(r)}+r^2d\Omega^2_2,
\label{m4}
\end{eqnarray}
where $\alpha(r)$ and $B(r)$ are functions of $r$. 
Of course the general formalism is also valid in the static case, and  leads 
to the horizon condition   
\begin{equation}
B(r_H)=0\,, \quad  B'(r_H)\neq 0\,,\quad e^{\alpha(r_H)} \neq 0\,.
\end{equation}
The Kodama vector reduces to
\beq
\mathcal K^\mu=\left(e^{-\alpha(r)}, \vec 0\right)\,.
\eeq
When $\alpha(r)=0$, which corresponds to case of GR in vacuum, 
the static Kodama vector coincides with the usual Killing vector 
$(1,\vec 0)$, and Hawking temperature of the related black hole reads
\begin{equation}  
T_K=\frac{1}{4\pi}\frac{d B(r_H)}{dr}\,.\label{T1}
\end{equation}
This is a well known result, and it can be justified in several ways, 
for example  making use of standard derivations of Hawking radiation \cite{V0}),
or by eliminating the conical singularity in the corresponding Euclidean metric, 
or making use of  the tunneling method, 
recently introduced in Refs.~\cite{PW,A,noi1}, and discussed in details in several papers.

However, as we shall see in explicit examples, within  modified gravity
it happens to deal with black hole solutions with $\alpha(r)\neq0$. 
In this case, the Kodama vector does not coincides with the Killing vector. 
Then one may introduce two Hawking temperatures, 
the Killing temperature (see, Appendix I)
\begin{equation}
T_K=\frac{1}{4\pi}\sqrt{\frac{d (e^{2\alpha(r)}B(r))}{d r}\frac{d B(r)}{d r}}\Big\vert_{r=r_H}
=\frac{1}{4\pi}e^{\alpha(r_H)}\frac{d B(r)}{d r}\Big\vert_{r=r_H}\,,
\end{equation}
and, making use of (\ref{H}) the Hayward temperature 
\be
T_H=\frac{\kappa_H}{2\pi}=\frac{1}{4\pi}\frac{d B(r)}{d r}\Big\vert_{r=r_H}\,,
\ee
which is trivially related to the previous one by $T_{K}=e^{\alpha(r_H)}T_H$.
If $\alpha(r)=0$ we recover Eq.(\ref{T1}), namely $T_K=T_H$. 
A detailed discussion about this issue can be 
found in Refs. \cite{noi2,noi3}, in which also the dynamical case is discussed. 

In the static case, all derivations of Hawking radiation 
(for example, the tunneling method in Appendix I) leads to a 
 semi-classical expression for the black hole radiation rate
\begin{equation}
\Gamma\equiv e^{-\frac{\Delta E_K}{T_K}}\,,
\label{rkill}
\end{equation}
in terms of the change $\Delta E_K$ of the Killing energy $E_K$ \cite{A}, 
but if one uses the Kodama energy $E_H$ for the emitted particle, one has  
\begin{equation}
\Gamma\equiv e^{-\frac{\Delta E_H}{T_H}}\,.
\label{rkoda}\end{equation}
From the Eqs. (\ref{rkill}) and (\ref{rkoda}) one arrives at the identity
\begin{equation}
\label{kk}
\frac{\Delta E_H}{T_H}=\frac{\Delta E_K}{T_K}\,,
\end{equation}
which may interpreted as the First Law of black hole thermodynamics
as soon as  $\Gamma \equiv e^{-\Delta S}$, with $S$ the 
entropy of the black hole itself. 
As a result, in the static case the two temperatures $T_K$ and $T_H$ are equivalent.

With regard to entropy of the black hole, it is well known that in GR the so called Area Law is satisfied, and we have
\begin{equation}  
S_W=\frac{\mathcal{A}_H}{4 G}\,.
\end{equation} 
In GR and in the static case, the First Law of black hole thermodynamics in vacuum reduces to
\beq
dE=T_{H} d S_W\,, 
\label{fl1}
\eeq
where $E$ is the Misner-Sharp energy evaluated on the horizon. 

Now we come to the key point of our proposal. For a generic modified gravity theories, for example the $F(R)$ models, where
$R$ is the Ricci curvature, it seams very difficult to define in a reasonable way the analogue of the local 
Misner-Sharp mass (see Ref.~\cite{Cai}). 
As we will see, an exception is the higher-dimensional Lovelock gravity \cite{lovelock}.

For this reason, in this paper, an attempt is made for obtaining an  expression 
of energy associated with  black holes 
solutions in higher order modified gravitational models. 
The proposal consists in the  identification of the black hole energy with a quantity proportional to the constant of integration, which appears in the explicit solution. 
The identification is 
achieved making use of derivation of the First Law of black hole thermodynamics 
from the equations of motion, evaluating in an 
independent way the related black hole entropy via Wald method \cite{wald} (see the Appendix II) and the Hawking temperature via the 
quantum mechanics in curved space-time, for example the  tunneling method \cite{PW} or other standard equivalent methods.
 
This approach is also supported by the results obtained in Refs. \cite{ram09,eli}, where, on quite general grounds, generalizing the Jacoboson results on GR
(see the seminal paper \cite{jacob}), the equations of a modified gravitational theories are shown to be equivalent to the First Law of black hole thermodynamics. As it is well known, this issue may be of high relevance in substantiating the idea that gravitation might be a manifestation of thermodynamics of quantum vacuum \cite{tanu0}. 

The paper is organized as follows. In Section {\bf 2}, the Lovelock gravity 
\cite{lovelock} is revisited, and the approach here proposed is shown to work for such a case. 
In Section {\bf 3}, the four dimensional modified gravity models of the $F(R)$ 
type are investigated, and the method proposed is 
applied to several cases in Sections {\bf 4} and {\bf 5}. 
In Section {\bf 6}  a new  topological black hole solution is discussed and 
the method is shown to work, as  in Section {\bf 7}, 
where the  conformal Weyl gravity black holes are considered.
Finally Section {\bf 8} contains the conclusions. In two Appendices, for the sake of completeness, the tunneling method and Wald entropy method are briefly discussed.

\section{ Lovelock Black Hole Solutions}
In this section, as  warm up, we review Lovelock theory with the related static 
and spherically symmetric black hole solutions. 
This theory is  a very interesting higher dimensional generalization of Einstein gravity. 
In general, by making use of higher order geometrical invariants in the action,
in the metric formalism for the field equations one obtains 
fourth order partial differential equations. 
However, as Lovelock had shown, one can obtain second order differential equation by
making use of higher dimensional extended Euler densities, 
the so called $m$-th order Lovelock terms defined by
\begin{eqnarray}
  {\cal L}_m = \frac{1}{2^m} 
  \delta^{\lambda_1 \sigma_1 \cdots 
\lambda_m \sigma_m}_{\rho_1 \kappa_1 \cdots \rho_m \kappa_m}
  R_{\lambda_1 \sigma_1}{}^{\rho_1 \kappa_1} 
\cdots  R_{\lambda_m \sigma_m}{}^{\rho_m \kappa_m}\ ,
\qquad m=1,2,3,...
\end{eqnarray}
where  $R_{\lambda \sigma}{}^{\rho \kappa}$ is the Riemann tensor in 
arbitrary $D$-dimensions and $\delta^{\lambda_1\sigma_1
   \cdots\lambda_m\sigma_m}_{\rho_1 \kappa_1 \cdots \rho_m \kappa_m}$ is the 
generalized totally antisymmetric Kronecker delta defined by  
\begin{eqnarray}
\delta^{\mu_1\mu_2\cdots \mu_p}_{\nu_1\nu_2\cdots\nu_p}={\rm det}
\left(
\begin{array}{cccc}
\delta^{\mu_1}_{\nu_1}&\delta^{\mu_1}_{\nu_2}&\cdots&\delta^{\mu_1}_{\nu_p}\\
\delta^{\mu_2}_{\nu_1}&\delta^{\mu_2}_{\nu_2}&\cdots&\delta^{\mu_2}_{\nu_p}\\
\vdots&\vdots&\ddots&\vdots\\
\delta^{\mu_p}_{\nu_1}&\delta^{\mu_p}_{\nu_2}&\cdots&\delta^{\mu_p}_{\nu_p}
\end{array}
\right)\ .
\nonumber
\end{eqnarray}
The action for Lovelock gravitational theory reads 
\begin{eqnarray}
I=\int d^Dx \sqrt{-g}\left[-2\Lambda+\sum_{m=1}^k\left\{\frac{a_m}{m}{\cal L}_m\right\}\right]\,,
\label{total_action}
\end{eqnarray}  
where we defined the maximum order $k\equiv [(D-1)/2]$ and  $a_m$ are arbitrary constants. 
Here $[z]$ represents the maximum integer satisfying $[z]\leq z$. 
Hereafter we set $a_1=1$.  

For such a kind of theory, the equations of motion in vacuum are second order quasi-linear 
partial differential equations in the metric tensor and  read   
\begin{eqnarray}
{\cal G}_{\mu}{}^{\nu}=0 ,
\label{}
\end{eqnarray}
the Lovelock tensor ${\cal G}_{\mu}{}^{\nu}$ being given by 
\begin{eqnarray}
{\cal G}_{\mu}{}^{\nu}=\Lambda \delta_{\mu}^{\nu}
-\sum_{m=1}^{k}\frac{1}{2^{m+1}}\frac{a_m}{m} 
	 \delta^{\nu \lambda_1 \sigma_1 \cdots \lambda_m \sigma_m}_{\mu \rho_1 \kappa_1 
\cdots \rho_m \kappa_m}
       R_{\lambda_1 \sigma_1}{}^{\rho_1 \kappa_1} 
\cdots  R_{\lambda_m \sigma_m}{}^{\rho_m \kappa_m}\,.
\label{EOM}
\end{eqnarray}
As we said in previous Section, we shall focus our attention on static, 
spherically symmetric solutions, thus we look for metric of the form
\begin{eqnarray}
ds^2=-B(r)dt^2+\frac{dr^2}{B(r)}+r^2d\Omega^2_n,
\label{metric_ansatz}
\end{eqnarray}
where $d\Omega^2_n$ is the metric of a $n$-dimensional sphere  $S^n$ ($n=D-2$),
Such kind of theories become quite interesting for $D>4$, 
the four-dimensional case being equivalent to Schwarzschild-de Sitter,
since ${\cal L}_1=R$ and ${\cal L}_2$ is equal to the Gauss-Bonnet quadratic term,
which in four-dimensions is a topological invariant.
 
A direct evaluation of field equations gives \cite{wheeler}
\begin{eqnarray}
&&{\cal G}_t^t={\cal G}_r^r=-\frac{n}{2r^n}\frac{d\left[r^{n+1}W(r)\right]}{dr}
\ ,\label{of}\\
&&{\cal G}_i^j=-\frac{1}{2r^{n-1}}\frac{d^2 
\left[r^{n+1}W(r)\right]}{d^2 r}\ ,
\end{eqnarray}
where $W$ is given by 
\begin{eqnarray}
W(r)=\sum_{m=2}^{k}\frac{\alpha_m}{m} [1-B(r)]^m r^{-2m}+[1-B(r)]r^{-2}
-\frac{2\Lambda}{n(n+1)}\,, 
\end{eqnarray}
with $	\alpha_m=a_m\prod_{p=1}^{2m-2}(n-p)$.

For example, for $D=4$, $k=1$, and so one has the Schwarzschild-de Sitter solution, while 
for $D=5$, $k=2$, there is one Lovelock non trivial term 
(the Gauss-Bonnet, which in five-dimensions is not a topological invariant) 
and one has the Boulware-Deser solution \cite{B}. 
For  higher dimensions one has an algebraic equation of increasing complexity, 
but, as we shall see in the following, 
for our purposes it will be not necessary to know explicitly the expression 
for the solution $B(r)$.    

For the static metric in (\ref{metric_ansatz}) one has the Killing vector 
$V^\mu=(1,\vec{ 0})$ and since 
\begin{equation}
\nabla_\nu{\cal G}_{\mu}^{\nu}=0\,,\qquad  {\cal G}_{\mu\nu}={\cal G}_{\nu\mu}\,,
\end{equation}
the vector $J_\mu={\cal G}_{\mu \nu}{}V^{\nu}$ 
is covariantly conserved and gives rise to a Killing conserved charge. 
This corresponds to the quasi-local generalized Misner-Sharp mass which reads
\begin{equation}
E(r)=-\frac{1}{8\pi G}\int_\Sigma d \Sigma_\mu J^\mu
   =\frac{nV(\Omega_n)}{16\pi G}\int_0^r d\rho \frac{d(\rho^{n+1}W)}{d \rho}
    =\frac{nV(\Omega_n)}{16\pi G}r^{n+1}W(r)\,,
\label{EMS}
\end{equation}
where $\Sigma$ is a spatial volume at fixed time,  $d\Sigma_\mu=(d\Sigma, \vec{0})$, and
assuming spherical horizons, 
$V(\Omega_n)=\frac{2\pi^{n/2+1/2}}{\Gamma(n/2+1/2)}$. 

In the absence of matter Eq.~(\ref{of}) can be integrated and one has
\begin{eqnarray}
r^{n+1}W(r)=C\,,
\label{solc}
\end{eqnarray}
$C$ being a constant of integration which we will show to be  related to the mass 
of the black hole. 
On shell, that is at the horizon $r=r_H$, $B(r_H)=0$,  
Eqs.~(\ref{EMS}) and (\ref{solc}) leads to 
\begin{equation}
E_K=\frac{nV(\Omega_n)}{16\pi G}C \,,
\label{E}
\end{equation}
Now let us show that a First Law of black hole thermodynamics holds true, with the ``energy'' 
of the black hole solution, namely the Killing charge  
obtained below,  proportional to constant of integration $C$.
In the case of Lovelock gravity the validity of the First Law of black hole thermodynamics  
has been investigated 
in many places (see for example \cite{meyer,tanu,maeda,cailove}).
For the static case we present a direct and simple proof.

First of all we introduce the horizon defined by the existence 
of the largest positive root $r_H$ of
\begin{equation}
B(r_H)=0\,, \quad  \frac{d B(r_H)}{d r}\neq 0\,.
\end{equation}
Then from Eq. (\ref{solc}) we have the identity
\begin{equation}
C=r_H^{n+1}W_H=\sum_{m=2}^{k}\frac{\alpha_m}{m} r_H^{n+1-2m}+ r_H^{n-1}-\frac{2\Lambda r_H^{n+1}}{n(n+1)}\,.
\label{*}
\end{equation}
On the other hand, taking the derivative with respect to $r$ of Eq. (\ref{solc}) and putting $r=r_H$, and making use 
again of   Eq.(\ref{solc}), we obtain
\begin{equation}
\sum_{m=2}^{k}\frac{\alpha_m(n+1-2m)}{m} r_H^{n+1-2m}+(n-1) r_H^{n-1}-\frac{2\Lambda r_H^{n+1}}{n}=
\frac{d B_H}{dr}\left(\sum_{m=2}^{k}\alpha_m r_H^{n+2-2m}+ r_H^{n}\right)\,.
\label{***}
\end{equation}
Now, let us compute the ``thermodynamical'' change of $C$ with respect 
to a small change of $r_H$.
>From  Eq.(\ref{solc}) one has
\begin{equation}
dC=\left(\sum_{m=2}^{k}\frac{\alpha_m(n+1-2m)}{m} r_H^{n-2m}+(n-1) r_H^{n-2}-
\frac{2\Lambda r_H^{n}}{n}\right)d r_H\,.
\label{f}
\end{equation}
Making use of Eq.(\ref{***})  this expression may be rewritten in the form
\begin{equation}
dC=\frac{d B_H}{d r}\left(\sum_{m=2}^{k}\alpha_m r_H^{n+1-2m}+ r_H^{n-1}\right) dr_H\,.
\label{fin}
\end{equation}
Let us interpret the r.h.s of the latter identity.  Here we are dealing with a 
static, spherically symmetric metric admitting a Killing vector.
If there is an event horizon located at $r_H$, then the Hawking temperature of the related 
black hole is given by  Eq.(\ref{T1}). 

Now, all thermodynamical quantities associated with these black holes solutions
can be computed by standard methods. In particular,  
the entropy can be calculated by the Wald method \cite{wald,V,F} 
or other methods if you like, and one has (see for example 
\cite{tanu,maeda,olea})
\begin{equation}  
S_W=\frac{2\pi V(\Omega_n)}{8\pi G}r_H^n\left(1
+n\sum_{m=2}^{k}\frac{\alpha_m}{n+2-2m} r_H^{2-2m}   \right)\,.
\label{we}
\end{equation} 
As a result, from Eqs. (\ref{E}), (\ref{f}), and (\ref{we}), 
one has the First Law of black hole thermodynamics for Lovelock gravity, that is
\begin{equation}  
T_K\,dS_W=dE_K\,.
\label{fl1bis}
\end{equation} 

We have shown that for a generic Lovelock gravity, 
the  First Law of black hole thermodynamics  holds and one can identify the energy of
a static, spherically symmetric black hole with the constant of integration 
and Killing conserved charge.

The generalization to topological Lovelock black holes has been investigated 
in \cite{caitop}, and again the  First Law of black hole thermodynamics  has been shown to hold.

\section{$F(R)$ four-dimensional modified gravity}

In this Section we will come back to $D=4$. 
To begin with, we recall  that the  action of modified $F(R)$-theories reads
\begin{equation}
I=\frac{1}{16 \pi G}\int d^4 x\sqrt{-g}\,F(R)\,,\label{action} 
\end{equation}
where $g$ is the determinant of metric tensor $g_{\mu\nu}$, and $F(R)$ is a generic function of the Ricci scalar $R$. For dimensional reason, $F(R)$ may contain a multiplicative functional dependence on $G$, the Newton constant. 

The equations of motion in vacuum for a general $F(R)$ model read
\begin{equation}
R_{\mu\nu}-\frac{1}{2}Rg_{\mu\nu}= G^{{\mathrm{MG}}}_{\mu\nu}
 \,.\label{EqEinstMod}
\end{equation}
Here, $R_{\mu\nu}$ is the Ricci tensor and the part of `modified gravity' ($MG$) is formally included into the tensor
$G^{{\mathrm{MG}}}_{\mu\nu}$, which is given by
\begin{equation}
G_{\mu\nu}^{{\mathrm{MG}}}=\frac{1}{F'(R)}\left\{\frac{1}{2}g_{\mu\nu}[F(R)-RF'(R)]
+(\nabla_{\mu}\nabla_{\nu}-g_{\mu\nu}\Box)F'(R)\right\}\,.\label{Tmod}
\end{equation}
The prime denotes derivative with respect to the curvature $R$,
$\nabla_{\mu}$
is the covariant derivative operator associated with $g_{\mu\nu}$ and
$\Box\phi\equiv g^{\mu\nu}\nabla_{\mu}\nabla_{\nu}\phi$ is the
D'Alembertian of a scalar field $\phi$. The trace of Eq.(\ref{EqEinstMod}) gives
\begin{equation}
3\Box F'(R)+RF'(R)-2F(R)=0\,, \label{scalaroneq}
\end{equation}
which shows that there exists an additive  scalar dynamical degree 
of freedom represented by $F'(R)$.

In the metric (\ref{m4}) the scalar curvature reads
\begin{eqnarray}
R  &=&
-3\,\left[{\frac{d}{dr}}B\left(r\right)\right]{\frac{d}{dr}}
\alpha\left(r\right)-2\,B\left(r\right)\left[{\frac{d}{dr}}
\alpha\left(r\right)\right]^{2}-{\frac{d^{2}}{d{r}^{2}}}
B\left(r\right)-2\,B\left(r\right){\frac{d^{2}}{d{r}^{2}}}\alpha\left(r\right)\nonumber\\
&&-4\,{\frac{{\frac{d}{dr}}B\left(r\right)}{r}}
-4\,{\frac{B\left(r\right){\frac{d}{dr}}\alpha\left(r\right)}{r}}-2\,{\frac{B\left(r\right)}{{r}^{2}}}
+\frac{2}{{r}^{2}}\,.\label{R}
\end{eqnarray}
In Ref.\cite{Zerbini} the following equations of motion have been found 
by Lagrangian methods \cite{vile,Capozziello,Monica}:
\begin{eqnarray}
\label{one}& &e^{\alpha(r)}\left(RF'(R)-F(R)-2F'(R)\frac{\left(1-B(r)-r(dB(r)/dr)\right)}{r^2}\right.\\\nonumber
& & \left.+2B(r)F''(R)\left[\frac{d^2 R}{d
r^2}+\left(\frac{2}{r}+\frac{dB(r)/dr}{2 B(r)}\right)\frac{d R }{d
r}+\frac{F'''(R)}{F''(R)}\left(\frac{d R}{d
r}\right)^2\right]\right)=0\,,
\label{2}
\end{eqnarray}
\begin{equation}
e^{\alpha(r)}\left[\frac{d\alpha(r)}{dr}\left(\frac{2}{r}+\frac{F''(R)}{F'(R)}\frac{d
R}{d r}\right)-\frac{F''(R)}{F'(R)}\frac{d^2 R}{d
r^2}-\frac{F'''(R)}{F'(R)}\left(\frac{d R}{d
r}\right)^2\right]=0\,.\label{two}
\end{equation}
Once $F(R)$ is given, together with Eq.(\ref{R}), the above equations form a system of three differential equations in the quantities $\alpha(r)$, $B(r)$ and $R(r).$ 

As already said, the static solutions describe a black hole 
if there exists a real positive solution $r_H$ of $B(r_H)=0,B'(r_H)\neq0$.
If this happens $r_H$ is the radius of the event horizon.
The Killing surface gravity reads
\begin{equation}
 \kappa_K \equiv\frac{1}{2}\sqrt{\frac{d (e^{2\alpha(r)}B(r))}{d r}\frac{d B(r)}{d r}}\Big\vert_{r=r_H}=\frac{1}{2}e^{\alpha(r_H)}\frac{d B(r)}{d r}\Big\vert_{r=r_H}\,.\label{k}
\end{equation}
Non trivial examples of such $F(R)$ gravity black hole solutions have been discussed 
in \cite{Zerbini} and we shall deal with them in the next Sections. 
In the following, we shall show that the  First Law of black hole thermodynamics holds, making use of the 
equations of motion and of the 
expressions for the Hawking temperature and entropy obtained by independent methods.
First, the tunneling method gives for the Hawking temperature 
\begin{equation}  
T_{K}=\frac{e^{\alpha(r_H)}}{4\pi}\frac{d B(r_H)}{dr}\,.
\end{equation}

The entropy associated to these black holes solutions
can be calculated by the Wald method (see Appendix II). One has
\begin{equation}  
S_W=\frac{\mathcal{A}_H}{4G}F'(R_H)\,.\label{wald}
\end{equation} 
For simplicity we will consider only spherical horizons, 
thus the area is $\mathcal A_H=4\pi r_H^2 $ and the volume
$V_H=\frac{4}{3}\pi r_H^3\,.$ 
By evaluating the equation of motion (\ref{one}) on the event horizon, 
and multiplying both sides of equation by $dr_H$, we have
\begin{equation}
T_{K}dS_W = e^{\alpha(r_H)}\left(\frac{F'_H}{2G}-\frac{R_HF'_H-F_H}{4 G}r_H^2\right)dr_H\,.
\label{capo}
\end{equation}
Thus, we have derived for a generic $F(R)$ gravitational model the
First Law of black hole thermodynamics  
as soon as the identification     
\begin{equation}
 E_{K}=\int\, e^{\alpha(r_H)}\left(\frac{F'_H}{2G}-\frac{R_HF'_H-F_H}{4 G}r_H^2 \right)dr_H
\,,\label{capo1}
\end{equation}
can be made.
Within  these $F(R)$ modified gravity theories, this is one of 
the main result of this paper. In the next Sections, 
by making use of several  exact solutions,  
we will provide a  support for this identification.

Our proposal, expressed by Eq. (\ref{capo1}), should  be compared with a
similar proposal contained in  Ref. \cite{cai0}. 
In Ref. \cite{cai1}  an attempt to define a local Misner-Sharp 
mass has been presented. There, however, the proposed formula is not really satisfactory, 
because the quasi-local form is only present in some particular cases, 
one of which will be discussed in the next Section.

\subsection{The constant curvature case} 

As a  simple but important example, let us consider the class of static solutions 
with constant curvature $R_0$. In this case one has the
solution with $\alpha=0$ (in \ref{m4}), 
and the the first equation of motion (\ref{one}) reduces to
\begin{equation}
\frac{d}{dr}\left(r-rB(r)+\frac{\Lambda_0\,r^3}{3}\right)=0\,,
\end{equation}
where
\begin{equation}
\Lambda_0=\frac{R_0F'_0-F_0}{2F'_0}\,.
\end{equation}
Thus, one arrives at Schwarzschild-de Sitter solution
\begin{equation}
 B(r)=\left(1-\frac{C}{r}-\Lambda_0\,\frac{r^2}{3}\right)\,, 
\end{equation}
and $R_0=4\Lambda_0$. Here $C$ is a constant of integration. 
The horizon is located at $r=r_H$, where 
\begin{equation}
1=\frac{C}{r_H}+\Lambda_0\frac{r_H^2}{3}\,.
\label{B0} 
\end{equation}
Making use of Eq. (\ref{capo1}) one has 
\begin{equation}
E_K= \frac{1}{2G}\left( F'_0(r_H-\frac{\Lambda_0 r_H^3}{3})\right)\,,
\end{equation}
and by Eq.(\ref{B0}) one finally gets
\begin{equation}
E_K=\frac{F'_0 C}{2G}\,, 
\end{equation}
which is our identification of  mass-energy expression for this class of black hole, 
in agreement with Ref.\cite{cai0}.

\section{The Clifton-Barrow solution}

Let us apply the same procedure for the highly non-trivial Clifton-Barrow solution\cite{CB}, 
for which $\alpha$ is not a constant. 
The  starting point is the following $F(R)$-modified gravity model:
\begin{equation}
F(R)=R^{\delta+1}G^\delta\,.
\end{equation}
For dimensional reasons we have also included the Newton constant $G^{\delta}$,
$\delta$ being a numerical parameter.
When $\delta=0$ the Hilbert-Einstein action of GR is recovered.
Note that in this case the modification with respect GR is not additive.

Looking for static, spherically symmetric metric of the type (\ref{m4}),
we find the Clifton-Barrow solution of Eqs.~(\ref{R})-(\ref{two}), that it
\begin{equation}
e^{\alpha(r)}=\left(\frac{r}{r_0}\right)^{\delta(1+2\delta)/(1-\delta)}\left(\frac{(1-2\delta+4\delta^2)(1-2\delta-2\delta^2)}{(1-\delta)^2}\right)^{1/2}\,,
\label{cb1}
\end{equation}
\begin{equation}
B(r)=\frac{(1-\delta)^2}{(1-2\delta+4\delta^2)(1-2\delta-2\delta^2)}\left(1-\frac{C}{r^{(1-2\delta+4\delta^2)/(1-\delta)}}\right)\,,
\end{equation}
and 
\begin{equation}
R=\frac{c_\delta}{r^2}\,,\qquad
c_\delta=\frac{6\delta(1+\delta)}{(2\delta^2+2\delta-1)}
\label{r}
\end{equation}
Above, $r_{0}>0$ is an arbitrary constant while $C>0$ is the 
integration constant of the model. We assume $\delta\neq1$.

The horizon radius, defined by $B(r_{H})=0$ and $\partial_{r}B(r_{H})\neq 0$ reads
\begin{equation}
r_{H}=C^{(1-\delta)/(1-2\delta+4\delta^2)}\,,
\label{c}
\end{equation}
and since $C>0$, the Clifton-Barrow metric is a black hole solution.

According to Equation (\ref{k}) the Killing-horizon surface gravity reads
\begin{equation}
\kappa_K=\frac{1}{2}\sqrt{\frac{(1-2\delta+4\delta^2)}{(1-2\delta-2\delta^2)}}\frac{r_{H}^{(2\delta+2\delta^2-1)/(1-\delta)}}{r_0^{\delta(1+2\delta)/(1-\delta)}}\,,
\end{equation}
which can be used to find the Killing-Hawking temperature $T_{K}=\kappa_K/2\pi$. 

With regard to the black hole entropy associated with the event horizon 
of the Clifton-Barrow solution, from the Wald 
formula in Equation (\ref{wald}) we find \cite{Bel}:
\begin{equation}
S_{W}=\frac{\mathcal{A}_H}{4 G^{1-\delta}}(1+\delta)\left[\frac{6\delta(1+\delta)}{(2\delta^2+2\delta-1)r_H^2}\right]^\delta\,. 
\end{equation}

In order to have the positive sign of entropy, we must require
$\delta>(\sqrt{3}-1)/2$ or $-1<\delta<0$. The solutions with
$0<\delta<(\sqrt{3}-1)/2$ or $\delta<-1$ are unphysical, whereas for
$\delta=0$ we find the result of General Relativity. On the other hand, only the solutions of $-1<\delta<0$ give a real value for the Killing surface gravity $\kappa_H$. If $\delta>(\sqrt{3}-1)/2$ the Hawking Temperature becomes imaginary.

Making use of Eqs.(\ref{capo}) one has
\begin{equation}
dE_K=A_\delta\, r_H^{(4\delta^2-\delta)/(1-\delta)}\,dr_H\,,
\qquad 
A_\delta=\frac14\,e^{\alpha_H}G^{\delta-1}(c_\delta)^\delta
         \left[2(1+\delta)-c_\delta\,\delta\right]\,.
\end{equation}
As a result, the energy  turns out to be
\begin{equation}
E_K=\frac{A_\delta(1-\delta)}{1+4\delta^2-\delta}\, r_H^{(1+4\delta^2-\delta)/(1-\delta)}\,. 
\end{equation}
Finally, from Eq. (\ref{c}) one gets again that the energy is proportional 
to the constant of integration of the BH solution since
\begin{equation}
 E_K=\frac{\Psi_\delta G^{\delta-1} }{r_0^{\delta(1+2\delta)/(1-\delta)}}\,C\,, 
\end{equation}
where we have introduced the dimensionless constant depending on $\delta$ 
\be
\Psi_\delta=\left(\frac{2^{\delta-1}3^{\delta}\delta^{\delta}(\delta-1)^2(\delta+1)^{\delta+1}}{\sqrt{1-2\delta-2\delta^2}\sqrt{1-2\delta+4\delta^2}}\frac{1}{(2\delta^2+2\delta-1)^{\delta}}\right)\,.
\ee

We conclude this Section with some remarks. In the above expression, the range of parameter $\delta$ has to be 
restricted to the ranges already discussed in order to have a positive temperature and entropy. As a check, it is easy 
to show that in the limit $\delta \rightarrow 0$, one gets the GR value  $C=2EG$. Furthermore, the Killing energy $E_K$ 
and the Killing temperature depend on the dimensional constant $r_0$, and we may take it proportional to Planck length 
$\sqrt G$.

\section{$1/R$ Model}

As a further non trivial example, let us consider the following $F(R)$-model:
\begin{equation}
F(R)=-\gamma\left(\frac{1}{R}-\frac{h^2}{6}\right), 
\end{equation}
where $h$ and $\gamma$ are positive, dimensional, arbitrary constants 
(we may choose, for example $\gamma=1/G^2$). 
In Ref.\cite{Zerbini} it has been shown that this model admits a
static, spherically symmetric solution of the type (\ref{m4})
\begin{equation}
e^{\alpha(r)}=\left(\frac{r}{r_0}\right)^{1/2}\,, 
\end{equation}
\begin{equation}
B(r)=\frac{4}{7}\left(1-\frac{7}{6h}r+\frac{C}{r^{7/2}}\right)\,, 
\end{equation}
and $R=6/(hr)$. Here $r_0$ is an arbitrary  constant, which is present 
for dimensional reasons, and $C$ is the integration constant. 
Let us consider the solution of $B(r_H)=0$, namely
\begin{equation}
\left(1-\frac{7}{6h}r_H+\frac{C}{r_H^{7/2}}\right)=0\,. 
\label{C}
\end{equation} 
If we assume $C>0$, it is easy to show that there exists always a simple zero 
$r_H>0$, which  defines  the event horizon, 
and so the above solution represents a black hole. 
With regard to the related entropy, Eq.(\ref{wald}) gives
\begin{equation}
S_{W}=\frac{\pi\gamma h^2 r_H^4}{36 G}\,,
\end{equation}
the entropy being positive, since $\gamma>0 $. 
The Killing temperature associates with the horizon reads
\begin{equation}
T_{K}=\frac{|\kappa_K|}{2\pi}=4
   \left(\frac{1}{6h}\sqrt{r_H}+\frac{C}{2r_H^{4}}\right)
    \left(\frac{1}{r_0}\right)^{1/2}\,. 
\end{equation}
By computing the Killing energy from  (\ref{capo1}) we have
\begin{equation}
E_K=\int  \frac{h\gamma}{54 G}\left(\frac{1}{r_0}\right)^{1/2}
     d\left(r_H^{9/2}-\frac{6}{7h} r_H^{7/2}\right)=
  \frac{h\gamma}{54 G}\left(\frac{1}{r_0}\right)^{1/2}\left(r_H^{9/2}
    -\frac{6}{7h} r_H^{7/2}\right)\,.
\end{equation}
Thus, making use of Eq.(\ref{C}) one arrives at 
\begin{equation}
E_K=\frac{h^2\gamma}{63 G}\left(\frac{1}{r_0}\right)^{1/2} C\,.
\end{equation}
Also in this case we can identify the integration constant of the model
 as a quantity proportional to the black hole Killing energy.

\section{The Deser-Sarioglu-Tekin topological black hole solutions}

In this Section, first we generalize the modified gravity black hole solution 
of Deser et al.~\cite{Deser:2007za}, and
then we shall show that also for these  solutions 
the  First Law of black hole thermodynamics is valid and the constant of integration  is proportional
to the Killing energy.  

For the sake of simplicity we shall restrict ourselves to the four-dimensional case, 
but, since we are interested in black hole with
generalized topological horizon, we have to include a non vanishing 
cosmological constant (see for example the GR case \cite{vanzo,altri2,Mann}). 
The $D$-dimensional case as well as the inclusion of Electromagnetism presents no difficulties.

To begin with, we write down the action of the model
\be
I = \frac{1}{16\pi} \int_{\mathcal M}\,d^4x\,\sqrt{-g} 
\left(R -2\Lambda + \sqrt{3}\sigma\,\sqrt{F}\right) \,, \label{d action}
\ee
where $\sigma$ is a real dimensionless parameter 
and $F=C_{\mu\nu\rho\delta}C^{\mu\nu\rho\delta}$ is the square of the Weyl tensor.  
For $\sigma=0$ the Weyl contribution turns off and GR result is recovered.
This model is a very interesting additive modification 
of GR with cosmological constant. 

For more generality we look for static, (pseudo)-spherically symmetric solutions 
with various topology and so we write the metric in the form
\be
ds^2 = -a^2(r)B(r) dt^2+\frac{dr^2}{B(r)}
     +r^2\,\left(\frac{d\rho^2}{1-k\rho^2}+\rho^2 d\phi^2\right)\,, \label{d metric}
\ee
where the horizon manifold will be a sphere $S_2$, a torus $T_2$ or a compact 
hyperbolic manifold $Y_2$, according to whether $k=1,0,-1$. 

A direct computation shows that the noteworthy properties
of the Weyl scalar $F$ discussed in Ref.\cite{Deser:2007za} for $k=1$, 
are still valid for $k=0,-1$. Thus the unknown functions $a(r)$ and $B(r)$ can be obtained
by imposing the stationary condition $\delta\hat I=0$, where, 
$\hat I$ is the original action evaluated on the metric (\ref{d metric}) 
(up to integration by parts and on the ``topological'' variable $\rho,\phi)$.
It reads
\be
\hat I=\int\,dr\,\left\{(1-\sigma)[r a'(r)B(r)+k a(r)] 
           + 3 \sigma a(r)B(r)-\Lambda r^2a(r)\right\}\,,
\ee
from which it follows
\be
(1-\sigma)ra'(r)+3\sigma a(r)= 0\,, 
\label{b}
\ee
\be
 rB'(r)+\frac{(1-4\sigma)}{1-\sigma}\,B(r) =k-\Lambda\,\frac{r^2}{1-\sigma} 
\label{a} \;.
\ee
Here we are  assuming  $\sigma\neq1,\frac{1}{2},\frac{1}{4}$. 
The general solutions are
\be
 a(r)= \left(\frac{r}{r_0}\right)^{\frac{3\sigma}{\sigma-1}}
\label{b1}\;,
\ee
\be
B(r)= k\,\frac{(1-\sigma)}{(1-4\sigma)} - Cr^{-\frac{1-4\sigma}{1-\sigma}}
          -\Lambda\,\frac{r^2}{3(1-2\sigma)}\,,
\ee
$C$  and $r_0$ being integration constants.

One can see that black hole solutions exists only for negative cosmological constant,
but in the case $k=1$, already discussed in \cite{Bel}, 
where $\Lambda$ can assume any arbitrary value.
As usual, the horizon is given by the positive root $r_H$ of $B(r)=0$ with $B'(r_H)\neq0$.
The algebraic equation can be easily solved for the integration constant $C$ 
and gives
\be
C=\left( k\,\frac{1-\sigma}{1-4\sigma} - \Lambda\,\frac{r_H^2}{3(1-2\sigma)}\right) 
      r_H^{\frac{1-4\sigma}{1-\sigma}}\,.
\label{c1}
\ee
The equation (\ref{a}) evaluated on  the horizon leads to 
\be
r_H B'(r_H)=k-\Lambda\,\frac{r_H^2}{1-\sigma}\,. 
\label{ah} 
\ee
Thus, since  the Killing-Hawking  temperature 
$T_K=a_HB'(r_H)/4\pi$, taking into account Eq. (\ref{b1}) we get
\be
4\pi r_H T_K=\left(k-\Lambda\,\frac{r_H^2}{(1-\sigma)}\right)\left(\frac{r_H}{r_0} \right)^{\frac{3\sigma}{\sigma-1}} \,.
\label{t1}
\ee 
On the other hand, a direct computation along the line discussed in \cite{Bel} 
for the case $k=1$ leads to the BH entropy
\be
S_W = \frac{\mathcal A_H}{4} \left(1 + \sigma\right) \;,
\label{result}
\ee
where, in order to deal with a positive entropy, 
we have to restrict to the interval $\sigma \in (-1,1)$. 
Above,  $\mathcal A_H=V_k r_H^2$, in which $V_1=4\pi$ (the sphere), 
$V_0=|\Im\,\tau|$, with $\tau$ the Teichmueller 
parameter for the torus, and finally 
$V_{-1}=4\pi g$, $g>2$, for the compact hyperbolic manifold with genus $g$ \cite{vanzo}. 

As a result we have
\be
 T_KdS_W=\frac{V_k \left(1 + \sigma\right)}{8\pi G}
\left(k-\Lambda\,\frac{r_H^2}{(1-\sigma)}\right)\left(\frac{r_H}{r_0} \right)^{\frac{3\sigma}{\sigma-1}}dr_H \,.
\label{t2}
\ee 
Furthermore, Eq. (\ref{c1}) gives
\be
dC=\left( k - \Lambda\,\frac{r_H^2}{1-\sigma}\right) r_H^{\frac{3\sigma}{\sigma-1}}dr_H\,.
\label{c1ultimo}
\ee
As a consequence the first Law holds and
\be
 E_K=\frac{V_k\left(1 + \sigma\right)}{8\pi G}C \,.
\label{t34} 
\ee 
In this class of modified gravitational models the energy of black hole is particularly simple, 
since the modification is describe by 
the dimensionless parameter $\sigma$.

\section{Topological Conformal Weyl Gravity}

In this Section, first we revisit the higher gravity black hole solution 
of Riegert and others \cite{r,mannah}, and its topological version \cite{klem}.

To begin with, we write down the action of the model in the form 
\be
I=\int_{\mathcal M} d^4 x\,\sqrt{-g}\,
  \left[ \gamma (R-2\Lambda)+3 \omega F\right] \,, \label{d action1}
\ee
where $\gamma$ is an arbitrary parameter, 
which may be proportional to the  square of Plank mass, 
$\omega$  is a dimensionless parameter  and 
$F=C_{\mu\nu\rho\delta} C^{\mu\nu\rho\delta}$ is the 
square of the Weyl  tensor. 
The pure conformal invariant model $\gamma=0$ is very interesting and its phenomenology 
has been investigated in Ref.\cite{m}.
 
As in previous Section, also here we shall consider 
various topology and this means that the metric will have the form (\ref{d metric}),
and the arbitrary functions $a(r),B(r)$ will be obtained from the reduced action
\be
\hat I=\int dr\,\left[\gamma\left(rB(r)a'(r)+ka(r) 
    -2\Lambda\,r^2a(r)\right)+\omega \frac{A^2(r)}{r^2a(r)}\right]\,, 
\ee
where we have put
\be
A(r)&=&r^2a(r)B''(r)+3r^2a'(r)B'(r)-2ra(r)B'(r)+2r^2a''(r)B(r)
      \nonumber\\&&\qquad
           -2ra'(r)B(r)+2a(r)B(r)-2ka(r)\,.
\ee
As a result, we are dealing  with a higher order Lagrangian system, 
the Lagrangian depending on the first and second 
derivative of the unknown functions  $a(r)$ and $B(r)$.

The equations of motion read
\begin{eqnarray}
&&4 \frac{d^2}{dr^2}\left(\frac{AB(r)}{a(r)}\right)
       -2\frac{d}{dr}\left(\frac{A}{ra(r)}[3rB'(r)-2B(r)]\right)
\nonumber\\ &&\qquad 
  +2\frac{A}{r^2a(r)}\left[r^2B''(r)-2rB'(r)+2B(r)-2k\right]
\nonumber\\ &&\qquad\qquad 
     -\frac{A^2}{r^2a^2(r)} 
      +\frac{\gamma}{\omega}\left[k-B(r)-rB'(r)-2\Lambda r^2\right]= 0\,, 
\label{bw} \end{eqnarray}
\begin{eqnarray}
&& \frac{d^2A(r)}{dr^2}-\frac{d}{dr}\left(\frac{A(r)}{ra(r)}[3ra'(r)-2a(r)]\right)
\nonumber\\ &&\qquad 
    +\frac{A(r)}{r^2a(r)}
       \left[2r^2a''(r)-2ra'(r)+2a(r)\right]+\frac{\gamma ra'(r)}{2\omega} = 0\,, 
\label{aw}
\end{eqnarray}
For simplicity let us look for exact solutions with $a(r)=1$. 
With this Ansatz Eq.~(\ref{aw}) can be integrated and one obtains
\be
B(r)=\frac{b_1}{r}+c_0+c_1r+c_2r^2\,,
\label{sol10}
\ee
$b_1$ and $c_k$ being integration constants. In order to satisfy Eq.~(\ref{bw})
we have to distinguish the two cases $\gamma\neq0$ (a modified Einstein gravity)
and $\gamma=0$, (pure conformal gravity),
since they provide completely different solutions.

In the case $\gamma\neq0$ Eq.~(\ref{bw}) is satisfied only if
\be
c_0=k\,, \quad c_1=c_2=0\,\quad c_3=-\frac13\Lambda\,,
\label{dsw}
\ee
while $b_1$ remains a free parameter.
We see that this is a topological Schwarzschild-de Sitter(AdS) black hole like solution,
since 
\be
B(r)=k-\frac{C}{r}-\frac13\,\Lambda r^2\,,
\label{ds}
\ee
where here $b_1$ has been replaced by $C$. It has to be noted that this is the
solution which one would have obtained from the Hilbert-Einstein action with
cosmological constant, that is with $\omega=0$. 

As we already said, if $\gamma=0$ the solution is completely different and in fact,
in such a case Eq.~(\ref{bw}) is satisfied only if
\be
c_1=\frac{c_0^2-k^2}{3b_1}\,.
\label{1}\ee
Now the solution depends on the three arbitrary parameters
$c_0$, $c_2$ and $b_1$. By a redefinition of them by 
$c_0\to k+3c_0$, $c_2\to\lambda$, $b_1\to-C$, we write it in the form
\be
B(r)=k+3c_0-\frac{c_0}{C}(2k+3c_0)\,r+\lambda r^2-\frac{C}{r}\,,
\label{kfin}
\ee
in agreement with the topological black hole solution already found by Klemm in \cite{klem}.

The event horizon exists as soon as there is positive solution $r_H$ of 
$B(r)=0$. For example, if $C>0$ and $\lambda=1/L^2>0$, 
it is easy to show that there exists always a positive root independently on the values 
of $c_0$ and of $L$, while, in the opposite case $\lambda<0$, 
a positive root of $B(r)=0$ exists only 
if $c_0\geq0$ and the value of $|\lambda|$ is sufficiently small.
The special $\lambda=0$ case will be discussed at the end of this Section.   

With regard to the computation of Entropy, assuming that there exists an event 
horizon $B(r_H)=0$, with $r_H>0$ and  $B'(r_H)\neq0$, 
for the pure Weyl gravity case the Wald method gives 
\be
S_W= 2\omega V_K\left(\frac{C}{r_H}-c_0\right)
= 2\omega V_K\left(x-c_0\right)
\quad\Longrightarrow\quad 
dS_W=2\omega V_K\,dx\,,
\label{en}
\ee
where for convenience we have introduced the variable $x=C/r_H$. 
Here $\mathcal A_H=V_kr_H^2$ ($k=1,0,-1$),
with $V_1=4\pi$, for the sphere, $V_0=|\Im\,\tau|$, $\tau$ being the Teichmueller 
parameter for the torus, 
and $V_{-1}=4\pi g$, $g>2$, for the compact hyperbolic manifold with genus 
$g$ \cite{vanzo}. 
The integration constant $C$ in Eq.~(\ref{en}) can be seen as a function of $r_H$ 
obtained by solving the equation $B(r_H)=0$, which, as it follows from (\ref{kfin}), 
it is a second-order algebraic equation in $C$. 
Of course, in order to have a positive entropy we have to choose $c_0<C/r_H=x$
and moreover $C$ has to be positive being proportional to the energy.

Now we restrict ourselves to the $\lambda=1/L^2>0$ case. In this way,
by solving the equation $B(r_H)=0$  with respect to $C$ we get
\be
2x=\frac{2C}{r_H}=\frac{r_H^2}{L^2}+k+3c_0+\sqrt W\,\,,
\quad W=\left(\frac{r_H^2}{L^2}+k+3c_0\right)^2-4c_0(2k+3c_0)>0\,,
\label{x}
\ee
and from the latter equation it follows
\be
dx=\frac{r_H}{L^2}\left(1+\frac{\frac{r_H^2}{L^2}+k+3c_0}{\sqrt W}\right)dr_H\,,
\quad
dC=r_Hdx+x\,dr_H\,.
\label{dx}
\ee
On the other hand the Hawking temperature can be written in the convenient form
\be
\label{dxx}
T_K=\frac{B'(r)\Big\vert_H}{4\pi}=\frac{1}{4\pi\,r_H}\left(\frac{2r_H^2}{L^2}+\sqrt W\right)\,,
\ee
and using Eqs.~(\ref{en}) and (\ref{dx}) we obtain
\be
T_KdS_W&=&\frac{\omega V_k}{2\pi L^2}\left(\frac{3r_H^2}{L^2}+k+3c_0+\sqrt W+
  \frac{2r_H^2}{L^2}\,\frac{\frac{r_H^2}{L^2}+k+3c_0}{\sqrt W} \right)dr_H 
\nonumber\\&& 
=\frac{\omega V_k}{\pi L^2}(r_Hdx+x\,dr_H)\,.
\label{tconf}
\ee 
We finally see that the First Law of black hole thermodynamics  reads
\be
 T_KdS_W= \frac{\omega V_k}{\pi L^2}dC\,.
\ee
As a result, we may again identify the energy as
\be
 E_K=\frac{\omega V_k}{\pi L^2}C \,.
\label{t3}
\ee 
We conclude this Section with some remarks. 
The pure Weyl conformal gravity does not contain dimensional parameters.
Thus, one could think that there exists a  trivial entropy and a vanishing energy,
but, as we have shown above, the solution gives rise to a length scale $L$
related to the  integration constant $\lambda$.
In such a case the  First Law of black hole thermodynamics holds 
and the energy of black hole solution 
is proportional to the other dimensional constant of integration $C$.

The situation is different when $\lambda=0$, since in such a case 
the scale does not emerge and for the horizon one gets
\be
\frac{r_H}{C}=\frac{k+3c_0+\sqrt{(c_0+k)(k-3c_0)}}{2c_0(2k+3c_0)}\,.
\ee
The latter equation gives a positive $r_H$ for $k\neq0$ and a suitable value for $c_0$.
In any case we see that $x=C/r_H$ is a pure number and so $dx=0$ and 
the entropy is trivially constant.
The  First Law of black hole thermodynamics  is trivially valid with a vanishing energy. 
This is the particular case discussed in \cite{klem}.

\section{Conclusions}

In this paper the issue of defining  the energy associated with  a 
static, spherically symmetric black hole 
solution in higher order modified gravitational models has been tackled. 
We have  proposed to  identify  the black hole energy as a quantity proportional 
to the constant of integration, 
which appears in the explicit black hole solution. 
The identification is substantiated by the fact that in all explicit and known examples, 
we have been able to show that the  First Law of black hole thermodynamics 
(Clausius relation) holds true as a consequence of equations of motion, 
and evaluating in an independent way the related entropy 
via Wald method  and the Killing-Hawking temperature via quantum mechanics techniques
in curved space time. 
In the case of $F(R)$ modified gravity 
some non trivial exact black hole solutions have been considered. 
In the case of modified gravity in which the quadratic Weyl scalar is additively present, 
first we have found the corresponding new topological black hole solution, 
and then we have verified for it our proposal. 
Finally our proposal has been shown to work also in another non trivial higher 
order gravity theory, namely  the topological conformal Weyl gravity. 
It is easy to show that the proposal is also working for constant curvature 
black holes solutions in the usual Einstein gravity with cosmological 
constant modified by generic curvature-squared terms (see, for example \cite{pope}).  

On general grounds, we may say that our explicit results are in agreement 
with the general result obtained recently in Ref. \cite{ram09}, 
and together other results appeared in literature seem to 
indicate that the thermodynamic origin of a generalized modified gravity, 
when horizons are present,  has a broad validity.

\section{Appendix I: the Tunneling method}

In this Appendix, for the sake of completeness, we present  a short review of the tunneling method in its Hamilton-Jacobi
variant \cite{noi1}. The method is based on the computation of the classical action $I$ along a trajectory starting slightly behind the trapping horizon but ending in the bulk, and the associated WKB approximation ($c=1$)
\beq
\mbox{Amplitude} \propto e^{i \frac{I}{\hbar}}\, .
\eeq 
The related semi-classical emission rate reads 
\beq
\Gamma \propto |\mbox{Amplitude}|^2 \propto 
e^{-2\frac{\Im\,I}{\hbar}} \,. 
\eeq
The imaginary part of the classical action is due to deformation of the integration path according to the Feynman 
prescription, in order to avoid the divergence present on the horizon. As a result, one asymptotically gets a 
Boltzmann factor, in which an energy $\omega$ appears, i.e.
\beq
\Gamma \propto e^{- \frac{\beta}{\hbar} \omega}\,,
\eeq
and the Hawking temperature is $T=\frac{1}{\beta}$.

To evaluate the action $I$, let us start with a generic 
static, spherically symmetric  solution 
in D-dimension, written in Eddington-Finkelstein 
gauge, which, as it is well known,  is regular gauge on the  horizon
\begin{eqnarray}
ds^2=-B(r)e^{2\alpha(r)}dv^2+2e^{\alpha(r)} dr\,dv}+r^2d\Omega^2_{D-2}=\gamma_{ij}(x^i)dx^idx^j+r^2d\Omega^2_{D-2\,.
\end{eqnarray}
here $x^i=(v,r)$, where $v$ is the advanced time. Since we are dealing with 
static, spherically symmetric  solution 
space-times, one may restrict to  radial trajectories, and only the two-dimensional normal metric is relevant, and the Hamilton-Jacobi equation for a (massless) particle is
 
\begin{eqnarray}
\gamma^{ij}\partial_i I \partial_j I=+2e^{\alpha(r)}\partial_v I \partial_r I+e^{2\alpha(r)}B(r)(\partial_r I)^2 =0\,.
\label{a2}
\end{eqnarray}
Thus
\begin{eqnarray}
 \partial_r I=\frac{2\omega}{e^{\alpha(r)}B(r)}\,.
\label{a3}
\end{eqnarray}
in which $\omega=-\partial_v I$ is the Killing energy of the emitted particle. In the near horizon approximation,
$B(r)\simeq B'_H(r-r_H)$. As a consequence, making use of Feynman prescription for the simple pole in $r-r_H$, one has
\begin{eqnarray}
I=\int dr \partial_r I=\int dr \frac{2\omega}{e^{\alpha(r)}B'_H(r-r_H-i\varepsilon)}\,,
\label{a4}
\end{eqnarray}
where the range of integration over $r$ contains the location of the horizon $r_H$. Thus
\begin{eqnarray}
\Im\,I= \frac{2\pi \omega}{e^{\alpha_H}B'_H}\,,
\label{a5}
\end{eqnarray}
and the Hawking-Killing temperature is
\begin{eqnarray}
T_K=\frac{e^{\alpha_H}B'_H}{4 \pi}\,.
\label{a6}
\end{eqnarray}
If one had introduced the Kodama energy $\omega_H=e^{-\alpha_H}\omega$, one would have obtained the Hayward temperature 
$T_H=\frac{B'_H}{4 \pi}\,$. 

\section{Appendix II: the Wald Entropy}

In this Appendix, we recall the basic formula for computing black hole entropy for a generalized covariant theory of gravity. Following \cite{wald,V,F}, the explicit calculation of the black hole entropy $S_W$ is provided by the formula
\beq
S_W = - 2\pi \int_{S} \left(\frac{\delta \mathscr L}{\delta R_{\alpha \beta \gamma \delta}}\right)_{H}\, 
e_{\alpha \beta} e_{\gamma \delta} d S\,,
\eeq
where $\mathscr L = \mathscr L(R_{\alpha \delta \gamma \delta}) $ is the Lagrangian density of any general theory of gravity, 
$e_{\alpha\beta}=-e_{\beta\alpha}$, 
is the binormal vector to the (bifurcate) horizon. 
It is normalized so that $e_{\alpha\beta}e^{\alpha\beta}=-2$. 

For the metric (\ref{m4}), the binormal turns out to be
\beq
e_{\alpha\beta}= e^{\alpha(r)}(\delta^0_\alpha \,\delta^1_\beta - \delta^1_\alpha \,\delta^0_\beta )\;.
\eeq
The induced area form, on the bifurcate surface $r=r_H$, $t=$constant, is represented by $dS$.
Finally, the subscript $(H)$ indicates that the partial derivative  is evaluated on the horizon, and the variation of the Lagrangian density with respect to $ R_{\alpha \beta \gamma \delta}  $ is performed as if $R_{\alpha \beta \gamma \delta} $ and 
the metric $g_{\alpha \beta}$ are independent. For example,
\beq
\frac{\delta R}{\delta R_{\mu \nu \alpha \beta }}=
\frac{1}{2}\left(g^{\alpha \mu}g^{\nu \beta}-g^{\nu \alpha}g^{\mu \beta}  \right)\,. 
\eeq
As a result, for the modified gravity model of the $F(R)$ class, one obtains
\beq
S_W=\frac{{\cal A}_H\,F'_H}{4G}\,.
\eeq
This is the formula used in Section III and IV. 

\section*{Acknowledgments}
We thank L. Vanzo for discussions. We also would like to thanks the referee for
several useful remarks that have permitted to improve the final version of the paper.

\end{document}